\documentstyle[preprint,seceq]{jpsj}
\title
{
High-Temperature Dynamics of Spin Glasses
}

\author
{ 
Michiko {\sc Yamana},~ Hidetoshi {\sc Nishimori},~ Tadashi {\sc Kadowaki}
and D. {\sc Sherrington}$^{1}$
}
 
\inst
{
Department of Physics, Tokyo Institute of Technology,\\
Oh-okayama, Meguro-ku, Tokyo 152\\
$^1$Department of Physics -- Theoretical Physics, 
University of Oxford,\\
1 Keble Road, Oxford OX1 3NP, United Kingdom
}

\recdate
{
1997
}

\def\runauthor{
Michiko {\sc Yamana} ,~Hidetoshi {\sc Nishimori} ,~ Tadashi {\sc Kadowaki}
and D. {\sc Sherrington}
}

\abst
{
We develop a systematic expansion method of physical
quantities for the SK model and the finite-dimensional
$\pm J$ model of spin glasses in non-equilibrium states.
The dynamical probability distribution function
is derived from the master equation using a high temperature expansion.
We calculate the expectation values of physical quantities
from the dynamical probability distribution function.
The theoretical curves show satisfactory agreement with Monte Carlo
simulation results in the appropriate temperature and
time regions.
A comparison is made with the results of a dynamics theory by
Coolen, Laughton and Sherrington.
}

\kword
{
dynamics, SK model, $\pm J$ model, dynamical probability distribution
function, high-temperature expansion, CLS theory
}

\begin{document}
\sloppy
\maketitle
%
\section{Introduction}

Dynamics plays essential roles in understanding experimental
observations on spin glasses.  It is very difficult, however,
to develop systematic theoretical methods to investigate
dynamical behavior of spin glass models, and numerical
approaches have been the major source of information
on dynamical properties of finite-ranged spin glasses
until recently.
Recent activities in analytical studies on the dynamics of the
Sherrington-Kirkpatrick (SK)\cite{ref:SK}
model of spin glasses include closed-form solutions 
for dynamical correlation functions \cite{ref:Cogliandoro}
and construction of evolution equations for single-time
macroscopic quantities.\cite{ref:CLS,ref:LCS}

In the latter approach, Coolen, Laughton and Sherrington (CLS)
derived a closed-form evolution equation for the 
averaged single-site
spin-field distribution function under a few assumptions on
microscopic properties of the system.  The averaged
spin-field distribution function thus
obtained provides sufficient
information to determine the time evolution of single-time
macroscopic quantities. It gives quite good agreement with
the evolution of macroscopic quantities obtained by
computer simulation of the microdynamics, although a
high-temperature expansion of the dynamical microscopic
probability distribution\cite{ref:NY} 
involved quantities not expressible
solely in terms of the spin-field distribution
of CLS.

The purpose of the present paper is to further develop the
high-temperature expansion technique\cite{ref:NY} and
derive explicit expressions of physical quantities to
the third order in the inverse temperature.  The obtained
series results are still too short to discuss critical properties
around the spin glass transition temperature.
However, our method has an advantage that it works
not only for the infinite-range SK model but also for 
finite-dimensional systems, in particular the 
nearest neighbor $\pm J$ model.
Thus the present approach is a first step toward a systematic
investigation of dynamics of finite-dimensional spin glasses.

In the next section the problem and the method are formulated.
The explicit expression of the dynamical probability distribution
function is derived there to the third order of the inverse temperature.
A method to evaluate expectation values of physical quantities
using the dynamical probability distribution function
is formulated in $\S 3$.
This formulation is used in $\S 4$ to determine the coefficients
of series expansions of several physical quantities.
In $\S 5$ the averaged spin-field distribution function is evaluated
by the expansion method.  The results in these two sections are
used in $\S 6$ to discuss the limit of applicability of the CLS theory.
The final section is devoted to general discussions.

\section{Dynamical probability distribution function}
%
\newcommand{\hh}{\sum_i h_i^2}
\newcommand{\sh}{\sum_i \sigma_i h_i^3}
\newcommand{\Jhh}{\sum_{ij} J_{ij} h_i h_j}
\newcommand{\J}{ \{ J_{ij} \} }
\newcommand{\sums}{\sum_{{\mib \sigma}}}
%
The essence of our theory is a high-temperature expansion of the 
microscopic probability distribution function.
The basic formulation of this expansion was developed in 
Ref.~\citen{ref:NY}
for the SK model of spin glasses.
We describe here briefly the method generalized to include 
finite-dimensional models.

The model we consider is an Ising spin glass with Hamiltonian
\begin{equation}
   H({\mib \sigma}) =-\sum_{\langle ij\rangle}J_{ij} \sigma_i\sigma_j ,
   \label{Hamiltonian_t}
\end{equation}
where ${\mib \sigma}$ denotes the spin configuration
and the $\{ J_{ij}\}$ are quenched randomly distributed;
$\langle ij\rangle$ denotes a pair of sites.
The probability distribution function $p_t({\mib \sigma})$
obeys the master equation:
\begin{equation}
   \frac{1}{p_t({\mib \sigma})}\frac{{\rm d}}{{\rm d}t}p_t({\mib \sigma})
   =\frac{1}{p_t({\mib \sigma})}\sum_k
     {p_t(F_k {\mib \sigma})}w_k(F_k{\mib \sigma})
     -\sum_kw_k({\mib \sigma}),
   \label{master}
\end{equation}
where $F_k$ is a single spin flip operator, $F_k\Phi ({\mib \sigma})
\equiv \Phi (\sigma_1,\cdots,-\sigma_k,\cdots,\sigma_N)$,
$N$ being the system size,
and we use a transition rate of the heat bath method
\begin{equation}
 w_k({\mib \sigma})=\frac{1}{2}\left\{
    1-\sigma_k\tanh \beta h_k({\mib \sigma})\right\} .
  \label{t-rate}
\end{equation}
Here $\beta$ is the inverse temperature and
$ h_k({\mib \sigma})$ denotes the local field on
site $k$.

Let us solve the master equation (\ref{master}) by a high-temperature
expansion in the form
\begin{equation}
 p_t({\mib \sigma})=\exp \left\{ \beta f_t({\mib \sigma})
   +\beta^2 g_t({\mib \sigma})+\beta^3 u_t({\mib \sigma}) 
   +\cdots \right\}.
   \label{expansion}
\end{equation}
Inserting eq.~(\ref{expansion}) into the master equation (\ref{master})
and expanding the result in powers of $\beta$,~we obtain differential
equations for the functions appearing in the exponent of eq.~(\ref{expansion}):
\begin{eqnarray}
  \frac{{\rm d}f_t}{{\rm d}t}&=&\frac{1}{2}\sum_k (\Delta_k f_t 
  +2 \sigma_k h_k)
  \label{df-dt}
\\
 \frac{{\rm d}g_t}{{\rm d}t}&=&\frac{1}{2} \sum_k \Bigl( 
\frac{1}{2}(\Delta_k f_t)^2+ \Delta_k g_t + \sigma_k h_k \Delta_k f_t \Bigr)
 \label{dg-dt}  
\\
 \frac{{\rm d}u_t}{{\rm d}t}&=&\frac{1}{2} \sum_k \biggl\{ \frac{1}{6}
 (\Delta_k f_t)^3+\Delta_k f_t \Delta_k g_t + \Delta_k u_t 
\nonumber
\\
& &{}+\sigma_k h_k \Bigl( \frac{1}{2} (\Delta_k f_t)^2+
  \Delta_k g_t \Bigr) -\frac{2}{3}\sigma_k h_k^3 \biggr\},
\label{du-dt}
\end{eqnarray}
where $\Delta_k f_t=f_t(F_k{\mib \sigma})-f_t({\mib\sigma})$.

We now restrict ourselves to the nearest neighbor $\pm J$ model 
on a finite-dimensional lattice.
The SK model is recovered by taking an appropriate limit
of infinite coordination number.
The system is assumed to be initially in equilibrium at the
inverse temperature $\beta_0$:
\begin{equation}
   p_{t=0}({\mib \sigma})=
   \exp \left[ -\beta_0H({\mib \sigma})\right].
   \label{initial-distr}
\end{equation}
Equation (\ref{df-dt}) is easily solved to yield
\begin{equation}
  f_t({\mib \sigma}) = a(t)H({\mib \sigma}) ,
  \label{f-general}
\end{equation}
where
\begin{equation}
  a(t)= \alpha e^{-2t} -1.
  \label{a(t)}
\end{equation}
The parameter $\alpha$ denotes $1-\beta_0/\beta $.
It is necessary that $\beta_0$ and $\beta$ are smaller than
$\beta_{\rm c}$, the inverse of the critical temperature, 
because the high-temperature expansion is valid only when
the system stays in the paramagnetic phase.

Using eq.~(\ref{f-general}), the next order differential equation
(\ref{dg-dt}) is written explicitly as
\begin{equation}
  \frac{{\rm d}g_t}{{\rm d}t} = \frac{1}{2}\sum_k\Delta_k g_t 
  +(\alpha^2e^{-4t}-\alpha e^{-2t}) \sum_k h_k^2 .
  \label{dg}
\end{equation}
This equation has the following solution
\begin{equation}
  g_t({\mib \sigma}) = b_1 (t)\sum_i h_i^2 +b_2 (t),
  \label{g-general}
\end{equation}
where
\begin{equation}
  b_1 (t)= -\frac{\alpha^2}{2} e^{-4t} 
  -\Bigl(\alpha t-\frac{\alpha^2}{2}\Bigr) e^{-2t}
  \label{b_1 (t)}
\end{equation}
and
\begin{equation}
  b_2 (t)
  =NzJ^2 \biggl\{
  \frac{\alpha^2}{4}e^{-4t}+ \Bigl( \alpha t 
            -\frac{\alpha^2}{2}+ \frac{ \alpha}{2} \Bigr)
   e^{-2t} + \frac{\alpha^2}{4} - \frac{\alpha}{2}
     \biggr\}
   \label{b2}
\end{equation}
with $z$ being the coordination number.
Since eq.~(\ref{b2}) is independent on the spin configuration,
this term does not affect the following argument and shall
be ignored hereafter.

Similarly, using eqs.~(\ref{f-general}), (\ref{a(t)}), (\ref{g-general})
and (\ref{b_1 (t)}) for $f_t$ and $g_t$,
eq.~({\ref{du-dt}) is rewritten as
\begin{eqnarray}
 \frac{{\rm d}u_t}{{\rm d}t} &=& \frac{1}{2}\sum_k\Delta_k u_t
 -4zJ^2 (2\alpha e^{-2t} -1) b_1 (t) H({\mib \sigma})
\nonumber
\\
& &{} +(\frac{2}{3} \alpha^3 e^{-6t} -\alpha^2 e^{-4t}) \sum_k\sigma_k h_k^3
 \nonumber
 \\
 & & {} -2(2\alpha e^{-2t} -1) b_1 (t)\sum_{k,l} J_{kl} h_k h_l.
 \label{du}
\end{eqnarray}
The solution of this equation is
\begin{equation}
  u_t({\mib \sigma}) = c_1 (t)H({\mib \sigma}) +c_2 (t)\sum_i\sigma_i h_i^3 
  +c_3 (t)\sum_{i,j} J_{ij} h_i h_j,
 \label{u-general}
\end{equation}
where the coefficients satisfy
\begin{eqnarray}
  \dot{c_1} (t) &=& -2c_1 (t) -(12z-8)J^2 c_2 (t) 
\nonumber
\\
 & &{} -4zJ^2 b_1 (t) (2\alpha e^{-2t} -1)
  \nonumber
  \\
  \dot{c_2} (t) &=& -4c_2 (t) + \frac{2}{3}\alpha^3 e^{-6t} -\alpha^2 e^{-4t}
\label{eqs}
  \\
  \dot{c_3} (t) &=& -2c_3 (t) -2b_1 (t) (2\alpha e^{-2t} -1).
  \nonumber
\end{eqnarray}
%
In deriving these equations we have assumed that there are
no triangular loops on the lattice in the sense
\[
 J_{kl} J_{lm} J_{mk}=0,
\]
where indices denote neighboring sites.
This assumption excludes several types of
lattices from our considerations
such as the triangular lattice.
Although it is possible to remove this simplifying
assumption, resulting complicated formulas do not
lead to new physics.

Equations (\ref{eqs}) are solved under the initial 
condition $c_1=c_2=c_3=0$ as
\begin{eqnarray}
&& c_1(t)=J^2 \biggl[-\bigl(2z- \frac{2}{3}\bigr) \alpha^3 e^{-6t} +
         \Bigl\{\bigl(4z- \frac{4}{3}\bigr) \alpha^3 
\nonumber
\\
& &{}-(4z-2) \alpha^2-(10z-4) \alpha^2 t \Bigr\} e^{-4t} 
+ \Bigl\{ 2z \alpha^2 t  
\nonumber\\
& &{} - 2z \alpha t^2 - \bigl(2z-\frac{2}{3}\bigr) \alpha^3
         +(4z-2) \alpha^2 \Bigr\} e^{-2t} \biggr] 
\label{c_1 (t)}
  \\
&&  c_2 (t) = -\frac{1}{3}\alpha^3 (e^{-6t} - e^{-4t}) -\alpha^2 t e^{-4t}
  \label{c_2 (t)}
  \\
&&  c_3 (t) = -\frac{1}{2}\alpha^3 e^{-6t} 
            +\bigl(\alpha^3 -\frac{1}{2}\alpha^2 -2\alpha^2 t\bigr) e^{-4t}
\nonumber
\\
& &{} + \bigl(\alpha^2 t -\alpha t^2 -\frac{1}{2}\alpha^3 
        +\frac{1}{2}\alpha^2\bigr) e^{-2t} .
  \label{c_3 (t)}
\end{eqnarray}
Since eqs.~(\ref{df-dt}), (\ref{dg-dt}) and (\ref{du-dt}) are 
first-order differential equations, the above
expressions (\ref{f-general}), (\ref{g-general}) and (\ref{u-general})
with coefficients obtained above represent the unique solution.
We have therefore obtained the dynamical probability distribution function 
to the third-order in $\beta$ for the 
$\pm J$ model as
\begin{eqnarray}
   p_t({\mib \sigma}) = \exp &\biggl[& \beta a(t) H({\mib \sigma}) 
            + \beta^2 b_1 (t) \sum_i h_i^2 
\nonumber
\\
& &{}+ \beta^3 \Bigl( c_1(t) H({\mib \sigma}) + 
         c_2 (t) \sum_i \sigma_i h_i^3 
\nonumber
\\
& &{}    + c_3 (t) \sum_{i,j} J_{ij} h_i h_j \Bigr) +
         \cdots \biggr] ,
   \label{result}
\end{eqnarray}
where $a (t), b_1 (t), c_1 (t), c_2 (t)$ and $c_3 (t)$ are given in
eqs.~(\ref{a(t)}), (\ref{b_1 (t)}) and (\ref{c_1 (t)}) - (\ref{c_3 (t)}).

We have considered the $\pm J$ model on a finite-dimensional
lattice with $z$ nearest neighbors.
The results derived previously for the SK model\cite{ref:NY}
are recovered by taking the limit $N \to \infty$ with $z=N-1$
and $J=\tilde{J} / \sqrt{N} ~(\tilde{J} \sim O(1) )$.

The above solution (\ref{result}) is invariant under the gauge
transformation $\sigma_i\to \sigma_i\tau_i, J_{ij}\to
J_{ij}\tau_i \tau_j$ $(\tau_i=\pm 1)$. This invariance
is naturally expected from gauge invariance of the master
equation (\ref{master}).  In general, any term in the expansion
can be expressed in terms of
$I_{ij}\equiv J_{ij}\sigma_i\sigma_j$ which is the basic building
block of all gauge invariant quantities.  
More explicitly, terms to the third order are written as
\begin{eqnarray}
&&H = -\sum_{\langle ij\rangle } I_{ij} \\
&& \sum_i h_i^2 = \sum_{ilm}I_{il}I_{im} \\
&& \sum_i \sigma_i h_i^3 =\sum_{ilmn}I_{il}I_{im}I_{in}\\
&& \sum_{ij}J_{ij}h_i h_j = \sum_{ijlm} I_{ij}I_{il}I_{jm},
\end{eqnarray}
where all the indices of the $I$'s stand for nearest neighbor
sites.  Higher-order terms should be able to be written similarly.

The series-expansion solution obtained in the present section
forms the basis of estimation of physical quantities in the
following sections.

\section{High-temperature expansion of physical quantities}
We now apply the series expansion solution for the probability
distribution function to the expectation value of a physical
quantity $O$:
\begin{equation}
 \bigl\langle O ({\mib \sigma} , \{ J_{ij} \}) \bigr\rangle_t
 \equiv  \frac{ \sum_{ {\mib \sigma} } O( {\mib \sigma} , \{ J_{ij} \})
       p_t({\mib \sigma}) }
         { \sum_{ {\mib \sigma} } p_t({\mib \sigma}) }.
\label{O-ave}
\end{equation}
The first-order solution (\ref{f-general}) with eq. (\ref{a(t)}) gives
\begin{equation}
 p_t({\mib \sigma}) = \exp \bigl[ -\beta (1-\alpha e^{-2t})
 H({\mib \sigma}) \bigr] \equiv 
 \exp [ -\beta_{\rm eff} H({\mib \sigma} ) ].
\label{1st}
\end{equation}
This equation shows that the dynamical expectation value (\ref{O-ave})
can be calculated using a time-dependent
effective inverse temperature $\beta_{\rm eff}$.
Therefore we may regard the system as being in equilibrium
at temperature $T/(1-\alpha e^{-2t})$ at any given time $t$.
The expectation value with respect to this effective Boltzmann factor
(\ref{1st}) will be denoted by $\langle \cdots \rangle_1$ 
in the following.

The higher-order terms in the series expansion are taken into account
by the expansion
\begin{eqnarray}
&&\exp \Bigl\{ \beta a(t)H
      + \beta^2b_1(t) \sum_i h_i^2+\beta^3 
(c_1(t)H+\cdots)\Bigr\}
\simeq 
\nonumber
\\
&& e^{-\beta_{\rm eff} H} \Bigl\{ 
1+\beta^2 b_1(t) \sum_i h_i^2+\beta^3
(c_1(t)H+\cdots) \Bigr\}.
\end{eqnarray}
The thermodynamic average of a physical quantity 
$O({\mib \sigma},\{ J_{ij} \})$ then becomes
\begin{full}
\begin{eqnarray}
&&  \bigl\langle O ({\mib \sigma} , \{ J_{ij} \}) \bigr\rangle_t
\simeq  \frac{ \bigl\langle O 
        +\beta^2 b_1(t)  O \sum_i h_i^2 
        +\beta^3 \bigl( c_1(t)  O H({\mib \sigma}) 
        +\cdots \bigr) \bigl\rangle_1 }
       { \bigl\langle 1+\beta^2 b_1(t) \sum_i h_i^2 +\beta^3 \bigl(
           c_1(t) H({\mib \sigma})  +\cdots \bigr) \bigr\rangle_1 } 
\nonumber
\\
&\simeq&  \bigl\langle O \bigr\rangle_1 +\beta^2 b_1(t)\Bigl(
\bigl\langle O \sum_i h_i^2 \bigr\rangle_1
-\bigl\langle O \bigr\rangle_1
\bigl\langle \sum_i h_i^2 \bigr\rangle_1 \Bigr)
 +\beta^3 \Bigl\{ c_1(t)\Bigl( \bigl\langle O H({\mib \sigma}) 
\bigr\rangle_1 -\bigl\langle O \bigr\rangle_1 
\bigl\langle H({\mib \sigma}) \bigr\rangle_1 \Bigr)
\nonumber
\\
&+& c_2(t)\Bigl( \bigl\langle O \sum_i \sigma_i h_i^3 
\bigr\rangle_1 -\bigl\langle O \bigr\rangle_1 
\bigl\langle \sum_i \sigma_i h_i^3 \bigr\rangle_1 \Bigr)
 +  c_3(t)\Bigl( \bigl\langle O \sum_{i,j} J_{ij} h_i h_j 
\bigr\rangle_1 -\bigl\langle O \bigr\rangle_1 
\bigl\langle \sum_{i,j}J_{ij} h_i h_j \bigr\rangle_1 \Bigr)\Bigr\}
\nonumber
\\
&+&\cdots .
\label{Ot_ave3}
\end{eqnarray}
\end{full}
The configurational average of this equation is expressed as
\begin{eqnarray}
& &{} \bigl[ \bigl\langle O \bigr\rangle_3 \bigr]
  = \bigl[ \bigl\langle O \bigr\rangle_1 \bigr]+
 \beta^2 b_1(t) \mbox{Cov}\bigl( O \sum_i h_i^2 \bigr)_1
\nonumber
\\
&+&\beta^3 \Bigl\{ c_1(t)\mbox{Cov}
\bigl( O H({\mib \sigma}) \bigr)_1
+c_2(t) \mbox{Cov}\bigl( O \sum_i \sigma_i h_i^3 \bigr)_1
\nonumber
\\
&+&c_3(t)\mbox{Cov}
\bigl( O \sum_{ij} J_{ij} h_i h_j \bigr)_1\Bigr\}.
\label{ave3}
\end{eqnarray}
Here $\langle O \rangle_n$ stands for the thermodynamic average
obtained by truncating the exponent of eq.~(\ref{expansion})
at the $n$th order in $\beta$,
and $\mbox{Cov}(AB)_1 \equiv \bigl[\langle AB \rangle_1 \bigr]
-\bigl[ \langle A \rangle_1 \langle B \rangle_1 \bigr]$
represents the covariance of $A$ and $B$ at the effective
inverse temperature $\beta_{\rm eff}$.

Generally it is quite difficult to evaluate the expectation
value $\bigl[ \langle O\rangle_1 \bigr]$ and covariances 
on the right-hand side of eq.~(\ref{ave3})
for the finite-dimensional $\pm J$ model.
An exception is the first-order term $\bigl[ \langle O\rangle_1 \bigr]$
evaluated for a gauge-invariant quantity $O$ on the Nishimori line
defined by $\tanh \beta J=2p-1$,
where $p$ is the probability for $J_{ij}$ to be positive.\cite{ref:Ni}
In the case of the SK model, we can derive explicit expressions for
all terms in eq.~(\ref{ave3}) by the replica method.

Therefore we have to introduce an approximation to proceed further
for the $\pm J$ model.
For this purpose, we first focus our attention on the equilibrium
expectation values of terms appearing in the expansion (\ref{ave3})
evaluated on the Nishimori line:
\begin{eqnarray}
& &{}\Bigl[ \bigl\langle H({\mib \sigma}) \bigr\rangle_{\rm e} \Bigr]
 = - \frac{1}{2} NJz \tanh \beta J
\label{Heq}
\\
& &{} \Bigl[ \bigl\langle \hh \bigr\rangle_{\rm e} \Bigr]
 = NJ^2 \left\{ z + z(z-1) (\tanh \beta J)^2 \right\}
\label{hheq}
\\
& &{} \Bigl[ \bigl\langle \sh \bigr\rangle_{\rm e} \Bigr]
  = NJ^3 \left\{ z(3z-2) \tanh \beta J \right.
\nonumber
\\
&& + \left. z(z-1)(z-2) (\tanh \beta J)^3 \right\}
\label{sheq}
\\
&& \Bigl[ \bigl\langle \Jhh \bigr\rangle_{\rm e} \Bigr]
 = NJ^3 \left\{  z(2z-1) \tanh \beta J \right.
\nonumber
\\
&& + \left. z(z-1)^2 (\tanh \beta J)^3 \right\},
\label{Jheq}
\end{eqnarray}
where $\langle\cdots\rangle_{\rm e}$ denotes the equilibrium expectation
value.
The first of these equations, eq.~(\ref{Heq}), indicates that the factor 
$\tanh \beta J$ can be replaced by the expectation value of the Hamiltonian.
For instance,
\begin{equation}
 \bigl[ \bigl\langle \hh \bigr\rangle_{\rm e} \bigr]
  = NJ^2z -2J(z-1) \tanh \beta J
    \bigl[ \bigl\langle H({\mib \sigma}) \bigr\rangle_{\rm e} \bigr].
\label{expec}
\end{equation}
In equilibrium, both sides of eq.~(\ref{expec})
are of the order of $N$,
and fluctuations around the expectation value are
of order $\sqrt{N}$.  Thus, given a typical
equilibrium spin configuration, the value of the quantity
\[
\sum_i h_i^2
\]
is almost certainly equal to that of
\[
NJ^2z -2J(z-1) (\tanh \beta J) H({\mib \sigma}).
\]
This argument does not apply to temperatures near 
the critical point where fluctuations are not negligible.

The situation should not change drastically if the system is
close to equilibrium if not in equilibrium precisely.
Thus the following replacement may be a reasonable
approximation if the system is not far from equilibrium,
\[
  \hh \longrightarrow  NJ^2z -2J(z-1) (\tanh \beta J) H({\mib \sigma}). 
\]
Further {\it a posteriori} justification
of this approximate replacement comes from agreement
of the resulting values with simulations as shown
in the following.

We therefore apply similar replacements to all terms in the 
exponent of eq. (\ref{result}) using the relations
(\ref{hheq}) to (\ref{Jheq}).   The result is
\begin{full}
\begin{eqnarray}
p_t({\mib \sigma})&\simeq& \exp \Bigl( \beta a(t) H({\mib \sigma})
+ \beta^2 b_1(t) \Bigl\{ NJ^2z-2J(z-1)
(\tanh \beta J) H({\mib \sigma})\Bigr\}
\nonumber
\\
&+&\beta^3 \Bigl\{ c_1(t) -2c_2(t)J^2  
\Bigl( 3z-2+(z-1)(z-2) (\tanh \beta J)^2 \Bigr)
 \nonumber 
\\
&-&2c_3(t) J^2 \bigl(2z-1+(z-1)^2 (\tanh \beta J)^2
\bigr) \Bigr\} H({\mib \sigma}) \Bigr) .
 \label{bee}
\end{eqnarray}
\end{full}
This approximation is valid for a system near
equilibrium on the Nishimori line.

This result (\ref{bee}) indicates that the
dynamical probability distribution is 
approximated by the equilibrium Boltzmann factor with effective inverse
temperature
\begin{eqnarray}
&& (\beta_{\rm eff}^{(3)})_{\pm J}=-\beta a(t)+2\beta^2b_1(t)J(z-1)
 \tanh \beta J  -\beta^3 \times
   \nonumber\\
&& 
\left\{ c_1(t)-2c_2(t)J^2(3z-2)-2c_3(t)J^2(2z-1) \right\}.
\label{b-eff-J}
\end{eqnarray}
In the case of the SK model, the thermal expectation values
on the left-hand side of eqs.~(\ref{Heq}) - (\ref{Jheq})
can be evaluated using the replica method. The corresponding
$\beta_{\rm eff}^{(3)}$ for the SK model is then obtained as
\begin{eqnarray}
& &{}(\beta_{\rm eff}^{(3)})_{\rm SK}=
-\beta a(t)+\beta^3 \left(2{\tilde J}^2 b_1(t)-c_1(t) \right.
\nonumber
\\
&+ & \left. 6{\tilde J}^2c_2(t)+4{\tilde J}^2c_3(t)\right),
\label{b-eff-SK}
\end{eqnarray}
where ${\tilde J}^2/N$ represents the variance of
distribution of $J_{ij}$.
The above formula for the SK model can also be derived 
from eq.~(\ref{b-eff-J}), which is valid on the Nishimori
line, by taking the limit
$N \to \infty$, $z=N-1$ and $J=\tilde{J}/\sqrt{N}$ 
and applying the approximation
$\tanh \beta J \simeq \beta J$ in the $\pm J$ model.
Note that equilibrium properties of the SK model
in the paramagnetic phase
are independent of $J_0$, the center of distribution of
$J_{ij}$.\cite{ref:SK}
Therefore the results valid on the Nishimori line,
eqs.~(\ref{Heq}) - (\ref{Jheq}),
which passes through the paramagnetic phase
toward the multicritical point, remain true in the
whole region of the paramagnetic phase of the SK model.
This is the reason why eq.~(\ref{b-eff-SK}) is valid in the
whole paramagnetic phase of the SK model
whereas eq.~(\ref{b-eff-J}) for the finite-dimensional
nearest-neighbor $\pm J$ Ising model can be used only on the
Nishimori line.

In the form exhibited in eq.~(\ref{ave3}) the direct expansion
is useful in practice only for the infinite-ranged SK
model or to the level of the first-order term.
Quantitative estimation of validity of these methods will be 
studied in the next section.

\section{Explicit evaluation of expansion terms}
We now apply the formulas derived in the previous section to the
following physical quantities, 
$H,~\sum_i h_i^2,~\sum_i \sigma_i h_i^3$ and $\sum_{ij} J_{ij} h_i h_j$.
These four quantities constitute the expansion of the dynamical
probability distribution function to the third order in $\beta$
as in eq.~(\ref{result}).
One of the purposes to choose these quantities 
for a test ground of the method developed in the
previous section is to estimate 
the order of magnitude of various terms in eq. (\ref{result})
for consistency check
of the theory of Coolen, Laughton and Sherrington~\cite{ref:CLS,ref:LCS} 
as discussed in $\S 6$.

The first-order approximation for the $\pm J$ model is
obtained by replacing $\beta$ by $\beta_{\rm eff}$ in eqs.~(\ref{Heq}),
(\ref{hheq}), (\ref{sheq}) and (\ref{Jheq}).
The corresponding expressions for the SK model are found from
those of the $\pm J$ model by taking the limit of infinite
dimensionality, or alternatively, by applying the replica method
as described in Appendix A.
The results are
\begin{eqnarray}
&& \Bigl[ \bigl\langle H({\mib \sigma}) \bigr\rangle_1 \Bigr]
=-\frac{1}{2}N{\tilde J}^2 \beta_{\rm eff}
\\
&& \Bigl[ \bigl\langle \sum_i h_i^2 \bigr\rangle_1 \Bigr]
=N({\tilde J}^2+{\tilde J}^4\beta_{\rm eff}^2)
\\
&& \Bigl[ \bigl\langle \sum_i \sigma_i h_i^3 \bigr\rangle_1 \Bigr]
=N(3{\tilde J}^4\beta_{\rm eff}+{\tilde J}^6 \beta_{\rm eff}^3)
\\
&& \Bigl[ \bigl\langle \sum_{ij} J_{ij} h_i h_j \bigr\rangle_1 \Bigr]
=N(2{\tilde J}^4\beta_{\rm eff}+{\tilde J}^6\beta_{\rm eff}^3).
\end{eqnarray}

The third-order formula (\ref{ave3}) can be explicitly evaluated
only for the SK model as mentioned before.
The necessary covariances are calculated in Appendix A using the replica
method:
\begin{eqnarray*}
& &{}\mbox{Cov}\bigl(  H^2 \bigr)=\frac{1}{2}N{\tilde J}^2
\\
& &{}\mbox{Cov}\Bigl(\bigl(\hh\bigr)^2 \Bigr)=N{\tilde J}^4(2+8{\tilde J}^2 \beta^2)
\\
& &{}\mbox{Cov}\Bigl(\bigl(\sh\bigr)^2 \Bigr)
 =N{\tilde J}^6(24+54{\tilde J}^2 \beta^2+18{\tilde J}^4 \beta^4)
\\
& &{}\mbox{Cov}\Bigl(\bigl(\Jhh\bigr)^2 \Bigr)
 =N{\tilde J}^6(10+24{\tilde J}^2 \beta^2 +26{\tilde J}^4 \beta^4)
\\
& &{}\mbox{Cov}\Bigl(H \hh \Bigr)=-2N{\tilde J}^4 \beta
\\
& &{}\mbox{Cov}\Bigl(H \sh \Bigr)=-N{\tilde J}^4(3+3{\tilde J}^2 \beta^2)
\\
& &{}\mbox{Cov}\Bigl(H \Jhh \Bigr)=-N{\tilde J}^4(2+3{\tilde J}^2 \beta^2)
\\
& &{}\mbox{Cov}\Bigl(\bigl(\hh\bigr)\bigl(\sh\bigr) \Bigr)
=N{\tilde J}^6 \beta (18+12{\tilde J}^2 \beta^2)
\\
& &{}\mbox{Cov}\Bigl(\bigl(\hh\bigr)\bigl(\Jhh\bigr) \Bigr)
=N{\tilde J}^6 \beta (12+12{\tilde J}^2 \beta^2)
\\
& &{}\mbox{Cov}\Bigl(\bigl(\sh\bigr)\bigl(\Jhh\bigr) \Bigr)
\\
&& =N{\tilde J}^6(12+42{\tilde J}^2 \beta^2+18 {\tilde J}^4 \beta^4).
\end{eqnarray*}
Inserting these results into eq.~(\ref{ave3}), we obtain
\begin{eqnarray*}
& &{} \Bigl[ \bigl\langle H({\mib \sigma}) \bigr\rangle_3 \Bigr]
= \Bigl[ \bigl\langle H({\mib \sigma}) \bigr\rangle_1 \Bigr]
-2N{\tilde J}^4b_1(t)\beta_{\rm eff}\beta^2
\\
&& ~~~+N \Bigl(\frac{1}{2}{\tilde J}^2c_1(t)
-3{\tilde J}^4c_2(t)-2{\tilde J}^4c_3(t)\Bigr) \beta^3
\\
& &{}\Bigl[ \bigl\langle \sum_i h_i^2 \bigr\rangle_3 \Bigr]
=\Bigl[ \bigl\langle \sum_i h_i^2 \bigr\rangle_1 \Bigr]+2N{\tilde J}^4 b_1(t)
\beta^2
\\
& &{}\Bigl[ \bigl\langle \sum_i \sigma_i h_i^3 \bigr\rangle_3 \Bigr]
=\Bigl[ \bigl\langle \sum_i \sigma_i h_i^3 \bigr\rangle_1 \Bigr]
+18N{\tilde J}^6 b_1(t)\beta_{\rm eff}\beta^2
\\
&& ~~~+N \bigl(-3{\tilde J}^4c_1(t)+24{\tilde J}^6c_2(t)
 +12{\tilde J}^6c_3(t)\bigr) \beta^3
\\
& &{}\Bigl[ \bigl\langle \sum_{ij} J_{ij} h_i h_j \bigr\rangle_3 \Bigr]
=\Bigl[ \bigl\langle \sum_{ij} J_{ij} h_i h_j \bigr\rangle_1 \Bigr]
+12N{\tilde J}^6b_1(t)
\\
&& \times\beta_{\rm eff}\beta^2
+N \bigl(-2{\tilde J}^4c_1(t)+12{\tilde J}^6c_2(t)
+10{\tilde J}^6c_3(t)\bigr) \beta^3.
\end{eqnarray*}
The approximation of a Boltzmann form with the inverse temperature,
eqs.~(\ref{b-eff-J}) and (\ref{b-eff-SK}),
is also used in the following analysis.

We have compared the theoretical curves 
obtained by the approximations
$\langle\cdots\rangle_1, \langle\cdots\rangle_3$ and
$\beta_{\rm eff}^{(3)}$ with each other and with 
Monte Carlo simulations for the SK model.
Figures 1 to 4 show the results for 
$H,~\hh,~\sh$ and $\Jhh$, respectively.
The initial temperature was $T_0=\infty$ and the final temperatures
were set to $T=2$ and $5$ in each figure.
The system size of simulations was $N=1000$ with $50$-sample averages
and standard deviations indicated.
The center of distribution is $J_0=0$, and the temperature
is expressed in units of $J/k_{\rm B}$ or ${\tilde J}/k_{\rm B}$.

For all quantities the theoretical curves follow simulations
faithfully at relatively high temperature, $T=5$.
When the temperature is lower, $T=2$, the first-order 
approximation deviates in the intermediate time region,
$t=1 \sim 2$, from simulations as well as from other theoretical curves.

We next investigate the nearest neighbor $\pm J$ model.
The straightforward expansion (\ref{ave3}) cannot be used in this
case because the covariances are difficult to evaluate explicitly.
We therefore compare the first-order term and the
$\beta_{\rm eff}^{(3)}$-approximation with simulations.

The results for four quantities of the $\pm J$ model on the square lattice
are shown in Figs. 5 to 8.  We have set $T_0=\infty$
and the final temperatures were $T=2$ and $5$.
Monte Carlo simulations were carried out for the system size
$N=50^2$ on the Nishimori line.
The plots show results averaged over $50$ samples
with standard deviations indicated.

The critical temperature is $T_{\rm c}\simeq 0.96$ for the two-dimensional 
$\pm J$ model on the Nishimori line,\cite{ref:ON} while the SK model
has $T_{\rm c}=1$.\cite{ref:SK}
In addition, the two-dimensional model has the problem of Griffiths
singularity at $T_{\rm G}=2.27$.\cite{ref:RSP}
Therefore it is not surprising that the $T=2$ results do not reproduce
simulation data very well while the $T=5$ case does.
The slightly better situation at $T=2$ for the SK model than in the $\pm J$
model may be related to the absence of Griffiths singularities in the
SK model.

\section{Spin-field distribution function}
We now calculate the dynamical single-site
spin-field distribution function by high-temperature series expansion.
This distribution function has more information than
individual physical quantities
in the sense that the physical quantities
such as $[\langle H({\mib \sigma}) \rangle ]$,
$[\langle \sum_i h_i^2\rangle ]$ and
$[\langle\sum_i \sigma_i h_i^3 \rangle ]$ can be
derived from the distribution function.
It should be noted that $[\langle \sum_{ij} J_{ij} h_i h_j \rangle ]$
cannot be calculated from the present distribution function,
because this distribution function does not carry
information on correlations
between fields at different sites.

The averaged single-site spin-field distribution function is defined by
\begin{equation}
\label{lfd1}
P(h,\sigma)=\left[\frac{1}{N}\biggl<\sum_{i}\delta_{\sigma \sigma_i}
                  \delta(h-h_i)\biggr>_t\right] ,
\end{equation}
where $\sigma =\pm 1$ should not be confused with the
microscopic spin configuration of the whole system ${\mib \sigma}$.
The equilibrium spin-field distribution has been calculated
by Laughton {\it et al.}\cite{ref:LCS}} as
\begin{equation}
\label{Laughton}
P(h,\sigma)=\frac{1}{2\sqrt{2\pi}{\tilde J}}
            \exp\left(-\frac{(h-\sigma {\tilde J}^2 \beta^2)}
{2{\tilde J}^2}\right) .
\end{equation}

Now we consider the dynamical single-site spin-field
distribution function.
This function should reduce to eq.~(\ref{Laughton})
in the equilibrium limit.
It is convenient to introduce the characteristic function
(Fourier transform):
\begin{equation}
\label{lfd2}
P(h,\sigma)=\int^{\infty}_{-\infty}\frac{d k}{2\pi}G(k,\sigma)\exp(-ikh) ,
\end{equation}
or
\begin{equation}
\label{lfd3}
G(k,\sigma)=\left[\frac{1}{N}\biggl<\sum_{i}\frac{1+\sigma\sigma_i}{2}
                  \exp(ik\sum_{j}J_{ij}\sigma_{j})\biggr>_t\right] .
\end{equation}

The method of high-temperature series expansion
developed in the previous sections is
applicable to the evaluation of eq.~(\ref{lfd3}).
Detailed calculations are described in Appendix B.
The result is
\begin{full}
\begin{eqnarray}
\label{lfd4}
& &{}G(k,\sigma) =  \frac{1}{2} \exp\left(-\frac{k^2{\tilde J}^2}{2}\right)
 \Bigr[\left\{\cos(k{\tilde J}^2\beta_{\rm eff})
            +i\sigma\sin(k{\tilde J}^2\beta_{\rm eff})\right\}
\nonumber\\
& +&\beta^2b_1\bigr\{-4k{\tilde J}^4\beta_{\rm eff}\sin(k{\tilde J}^2\beta_{\rm eff})
  - k^2{\tilde J}^4\cos(k{\tilde J}^2\beta_{\rm eff})
 + i\sigma\left(4k{\tilde J}^4\beta_{\rm eff}\cos(k{\tilde J}^2\beta_{\rm eff})
                  -k^2{\tilde J}^4\sin(k{\tilde J}^2\beta_{\rm eff})\right)\bigl\}
\nonumber\\
&+&\beta^3\Bigl\{c_1\bigr\{k{\tilde J}^2\sin(k{\tilde J}^2\beta_{\rm eff})
 +i\sigma\left(-k{\tilde J}^2\cos(k{\tilde J}^2\beta_{\rm eff})\right)\bigl\}
+c_2\bigr\{-3k^2{\tilde J}^6\beta_{\rm eff}\cos(k{\tilde J}^2\beta_{\rm eff})
\nonumber\\
           &-&({\tilde J}^6(6k\beta_{\rm eff}^2-k^3)+6k{\tilde J}^4)\sin(k{\tilde J}^2\beta_{\rm eff})
\nonumber\\
&+& i\sigma\left(-3k^2{\tilde J}^6\beta_{\rm eff}\sin(k{\tilde J}^2\beta_{\rm eff})
+({\tilde J}^6(6k\beta_{\rm eff}^2-k^3)+6k{\tilde J}^4)\cos(k{\tilde J}^2\beta_{\rm eff})\right)\bigl\}
+  c_3\bigr\{-2k^2{\tilde J}^6\beta_{\rm eff}\cos(k{\tilde J}^2\beta_{\rm eff})
\nonumber\\ 
&-&(6k{\tilde J}^6\beta_{\rm eff}^2+4k{\tilde J}^4)\sin(k{\tilde J}^2\beta_{\rm eff})
  \nonumber\\
      &+&i\sigma\left(-2k^2{\tilde J}^6\beta_{\rm eff}\sin(k{\tilde J}^2\beta_{\rm eff})
  +(6k{\tilde J}^6\beta_{\rm eff}^2+4k{\tilde J}^4)\cos(k{\tilde J}^2\beta_{\rm eff})\right)\bigr\}\Bigr\}\Bigl] .
\end{eqnarray}
\end{full}
Substituting eq.~(\ref{lfd4}) into eq.~(\ref{lfd2}),
we arrive at the expression of the spin-field distribution:
\begin{full}
\begin{eqnarray}
\label{lfd5}
& & P(h,\sigma) =  \frac{1}{2\sqrt{2\pi}{\tilde J}} \Biggl[ 1 
     + \beta^2 b_1 \Bigl\{ 4\sigma {\tilde J}^2\beta_{\rm eff}(h-\sigma {\tilde J}^2\beta_{\rm eff}) 
 - ({\tilde J}^2-(h-\sigma {\tilde J}^2\beta_{\rm eff})^2) \Bigr\} 
 \nonumber\\
    &+& \beta^3 \biggl\{ c_1 \Bigl\{ -\sigma(h-\sigma {\tilde J}^2\beta_{\rm eff}) \Bigr\}
\nonumber\\
&+&  c_2 \Bigl\{ -3{\tilde J}^2\beta_{\rm eff}({\tilde J}^2-(h-\sigma {\tilde J}^2\beta_{\rm eff})^2 
    -3\sigma {\tilde J}^2(h-\sigma {\tilde J}^2\beta_{\rm eff}) 
 + \sigma (h-\sigma {\tilde J}^2\beta_{\rm eff})^3 
  \nonumber\\
       &+&6\sigma {\tilde J}^2(1+{\tilde J}^2\beta_{\rm eff}^2)(h-\sigma {\tilde J}^2\beta_{\rm eff}) \Bigr\} 
\nonumber\\
&+& c_3 \Bigl\{ -2{\tilde J}^2\beta_{\rm eff}({\tilde J}^2-(h-\sigma {\tilde J}^2\beta_{\rm eff})^2 
    +6\sigma {\tilde J}^2(2+3{\tilde J}^2\beta_{\rm eff}^2)
 (h-\sigma {\tilde J}^2\beta_{\rm eff}) \Bigr\}\biggr\} \Biggr] 
   \nonumber\\
 &\times &
    \exp\left(-\frac{(h-\sigma {\tilde J}^2\beta_{\rm eff})^2}{2{\tilde J}^2}\right) .
\end{eqnarray}
\end{full}

It is straightforward to check that the expansion results of
$[\langle H({\mib \sigma}) \rangle ]$, $[\langle \sum_i h_i^2 \rangle ]$
and $[\langle \sum_i \sigma_i h_i^3 \rangle ]$
given in \S 4 are recovered from eq.~(\ref{lfd5})
by appropriate integrations.

In equilibrium the distribution (\ref{lfd5}) becomes
Gaussian.  However it is not Gaussian in general even
in the paramagnetic phase.
As the temperature is lowered,
the distribution deviates from Gaussian more significantly.

The results of the analytical calculation and computer simulations
are compared in Fig. 9.
Simulations were performed for $N=5000$ with 100 samples at $T=2$.
We may conclude that the third-order result (\ref{lfd5}) agrees
with simulations in a rather satisfactory manner at this temperature.

\section{Discussions on the CLS theory}
Coolen, Laughton and Sherrington (CLS)~\cite{ref:CLS,ref:LCS}
developed a theory of dynamics of the SK model.
They derived the closed-form evolution equation of a physical quantity
which determines macroscopic behavior of the system.
An important assumption in their derivation was that the average 
of a macroscopic physical quantity can be calculated under the ansatz
of equipartitioning of the dynamical probability distribution function.
Here equipartitioning means that the probability distribution function
is constant once the value of the single-site spin-field distribution function
$P(h,\sigma)$ is given.

The existence of the correlation term of fields at different sites,
$\Jhh$, in the probability distribution function (\ref{result})
shows that the equipartitioning ansatz of CLS
is not generally true and suggests that the CLS theory is an approximate
one.\cite{ref:NY}
We investigate in the present section the effects of the 
field-correlation term on physical quantities to determine the degree of
applicability of the CLS theory.

Figure 10 shows the third-order series-expansion curves 
and simulation results, superimposed for comparison, of various 
physical quantities.
This figure indicates that the field-correlation term is not necessarily
small compared with other terms appearing in the series expansion.
The time evolution of the correlation coefficients of 
$\Jhh$ and the other quantities is shown in Fig. 11 
obtained by the $\beta_{\rm eff}^{(3)}$-approximation.
The correlation coefficient between $A({\mib \sigma}, \{ J_{ij} \})$
and $\Jhh$ is defined as
\begin{eqnarray}
& &{}\mbox{Corr}\Bigl(A({\mib \sigma}, \{ J_{ij} \}) \bigl(\Jhh\bigr) \Bigr)
\nonumber
\\
&=& \frac{\mbox{Cov} \Bigl(A \bigl(\Jhh\bigr) \Bigr)}
{\sqrt{\mbox{Cov} \Bigl(A^2 \Bigr)}
 \sqrt{\mbox{Cov} \Bigl(\bigl(\Jhh\bigr)^2 \Bigr)}}.
\end{eqnarray}
Figure 11 shows that $\Jhh$ is closely correlated with $H$ and
$\sh$ but is not with $\hh$ in the initial time region.
Since the absolute value of correlation coefficients with $H$ and 
$\sh$ is close to unity, one may suppose that the field-correlation 
term $\Jhh$ is approximately replaced by $H$ or $\sh$,
which may be taken as a support for the CLS theory as an
approximation.
We can also obtain results similar to 
Fig. 10 for the nearest neighbor $\pm J$ model.

It may be useful to see the behavior of various terms in the 
series expansion with the time-dependent coefficients
$a(t),b_1(t),\cdots, c_3(t)$ taken into account as appearing in 
eq.~(\ref{result}).
Third-order series estimates are given for the SK model in 
Figs. 12 and 13 and in Fig. 15 
for the two-dimensional $\pm J$ model in the high temperature region.
Simulation results run almost along the same curves and are omitted 
in these figures.
Only simulation results are shown at lower temperatures,
$T=1$ for the SK model and $T=2$ for the $\pm J$ model,
because the expansion breaks down at these temperatures;
see  Figs. 14 and 16.

Our observation is that the field-correlation term multiplied by the 
coefficient, $\beta^3 c_3(t) \Jhh$, is very small compared to first 
and second order terms in the high-temperature region, 
$T=5$, both for the  SK and $\pm J$ models.
However, this is not necessarily the case in intermediate
time regions at lower temperatures; see Figs. 14 and 16
which show effects around $(t \sim 2)$.
Naively one might therefore expect to need to choose
appropriate parameter regions when applying the CLS theory.
However, trends of strong correlations for $\Jhh$ with $H$
and $\sum_i \sigma_i h_i^3$
as seen in Fig. 11 allow $\Jhh$ to be effectively
approximated by $H$ or $\sum_i \sigma_i h_i^3$, giving a support
for the CLS theory as an approximation
irrespective of the size of the term $\Jhh$.

It is useful to compare the macroscopic
spin-field distribution $P(h,\sigma)$
obtained by the CLS theory and that by the high-temperature
expansion as shown in Figs. 17 and 18.
Figure 17 shows the first-order cumulant, or the average,
of $P(h,1), P(h,-1)$ and $P(h,1)+P(h,-1)$ at $T=2$.
The simulation data are also displayed which do not seem to give a clear 
advantage either to the CLS theory or to the high-temperature
expansion.
The second-order cumulants are shown in
Fig. 18. 
The upper set of curves are for $P(h,1)+P(h,-1)$ and
the lower curves correspond to $P(h,1)$ and $P(h,-1)$.
This figure indicates that the simulation results
follow the CLS theory 
more closely than the high-temperature expansion
in this temperature region.
Therefore the CLS theory serves as quite a good approximation
for the macroscopic spin-field distribution function.
%
\section{Discussions}
We have derived explicitly the time dependence of macroscopic physical
quantities for the SK model and the nearest neighbor $\pm J$ model
on the Nishimori line.
Our formulation gives the microscopic dynamical probability distribution
function in the form of a high-temperature series expansion.
Evaluation of physical quantities has then been reduced to calculations
of various expectation values in equilibrium.  Such expectation values
can be calculated to any order in the inverse temperature $\beta$
in the case of the SK model.
For the $\pm J$ model on a finite-dimensional lattice, the situation
is more complicated and explicit evaluation is possible only
to the first order in $\beta$ on the Nishimori line.
We therefore introduced an approximation in which the dynamical
probability distribution has the same form as the equilibrium
Boltzmann factor with a time-dependent effective temperature.

The resulting expressions show excellent agreement with numerical
simulations at high temperatures.  Deviations are observed at lower
temperatures in the intermediate time region.  We anyway think it
a significant progress that a systematic method to evaluate explicitly
the time dependence of physical quantities has been formulated.

We have also analyzed the averaged spin-field distribution functions
obtained by the high-temperature expansion and the CLS theory
for the SK model.
Cumulants of the distribution functions show that the CLS theory
gives excellent agreement with simulations at a relative low
temperature where the high-temperature expansion is not very
useful.  Therefore the CLS theory, if not exact, serves as 
a reliable tool to analyze dynamics of the SK model.
%
\newcommand{\sia}{\sigma_i^\alpha}
\newcommand{\sja}{\sigma_j^\alpha}
\newcommand{\saa}{\sigma_a^\alpha}
\newcommand{\sba}{\sigma_b^\alpha}
\newcommand{\sca}{\sigma_c^\alpha}
\newcommand{\sda}{\sigma_d^\alpha}
\newcommand{\sea}{\sigma_e^\alpha}
\newcommand{\sib}{\sigma_i^\beta}
\newcommand{\sjb}{\sigma_j^\beta}
\newcommand{\sab}{\sigma_a^\beta}
\newcommand{\sbb}{\sigma_b^\beta}
\newcommand{\sag}{\sigma_a^\gamma}
\newcommand{\sbg}{\sigma_b^\gamma}
\newcommand{\scg}{\sigma_c^\gamma}
\newcommand{\sad}{\sigma_a^\delta}
\newcommand{\sbd}{\sigma_b^\delta}
\newcommand{\scd}{\sigma_c^\delta}
\newcommand{\sae}{\sigma_a^\epsilon}
\newcommand{\sbe}{\sigma_b^\epsilon}
\newcommand{\sde}{\sigma_d^\epsilon}
\newcommand{\see}{\sigma_e^\epsilon}
\newcommand{\saz}{\sigma_a^\zeta}
\newcommand{\sbz}{\sigma_b^\zeta}
\newcommand{\eX}{\exp \Bigl( \beta \sum_{i<j} J_{ij}\sum_\alpha
                 \sia \sja \Bigr) }
\newcommand{\eXd}{\exp \Bigl( \beta \sum_{i<j}{'} J_{ij}\sum_\alpha
                 \sia \sja \Bigr) }
\newcommand{\eXab}{\exp \Bigl( \beta J_{ab}\sum_\alpha
                 \saa \sba \Bigr) }
\newcommand{\eY}{\exp \biggl\{ \frac{{\tilde J}^2 \beta^2}{2N}
                 \sum_{i<j} \Bigl( \sum_\alpha \sia \sja \Bigr)^2 \biggr\} }
\newcommand{\eYab}{\exp \biggl\{ \frac{{\tilde J}^2 \beta^2}{2N} 
                  \Bigl( \sum_\alpha \saa \sba \Bigr)^2 \biggr\} }
\newcommand{\eYd}{\exp \biggl\{ \frac{{\tilde J}^2 \beta^2}{2N}
                  \sum_{i<j} {'} \Bigl( \sum_\alpha \sia \sja \Bigr)^2
                 \biggr\} }
\newcommand{\q}{q_{\alpha \beta}}
\newcommand{\eZ}{\exp \Bigl( \frac{{\tilde J}^2 \beta^2 N}{4}n \Bigr)}
\newcommand{\eZq}{\exp \Bigl( -\frac{{\tilde J}^2 \beta^2}{2} N 
 \sum_{\alpha < \beta} \q^2 + N \ln \sum_{{\mib \sigma}} e^L \Bigr) } 
\newcommand{\dq}{\Bigl\{ \prod_{\alpha < \beta} \sqrt{ \frac{N}{2\pi} }
            J \beta d\q \Bigr\} }
\newcommand{\sumt}{\sum_{ \{  \tau_{ij}  \} }}
\newcommand{\ptf}{\frac{\exp(K_p \tau_{ij})}{2\cosh K_p}}
\newcommand{\ebj}{\exp \Bigl( \beta J \sum_{\ij} 
\tau_{ij} \sigma_i \sigma_j \Bigr)}
\newcommand{\kazu}{\frac{1}{2}zN}
\newcommand{\ekp}{\exp \Bigl( K_p \sum_{\ij}  \tau_{ij}  \Bigr)}
\newcommand{\ckp}{(2\cosh K_p)}
\newcommand{\ij}{\langle ij \rangle}
\newcommand{\ta}{\tanh \beta J}
\section*{Acknowledgements}
This work was commenced while DS was at the Center for 
Nonlinear Studies, Los Alamos National Laboratory, Los Alamos, 
NM 87545, USA and he would like to thank the US
Department of Energy for support.  He would also like to 
thank the Tokyo Institute of Technology and the Sasakawa Fund 
for support for a vist to Japan.

\appendix
\section{Covariances}
We show in this Appendix the procedure to calculate the covariance between
two physical quantities $A({\mib \sigma}, \J)$ and $B({\mib \sigma}, \J)$
for the SK model defined as
\[
\mbox{Cov}(AB) \equiv \bigl[\bigl\langle AB \bigr\rangle \bigr]
-\bigl[ \bigl\langle A \bigr\rangle \bigl\langle B \bigr\rangle \bigr].
\]
Here $\langle~~\rangle$ is the thermal average and $[~~]$
represents the configurational average.
The exchange interaction obeys the Gaussian distribution 
with vanishing mean and variance ${\tilde J}^2/N$, 
\[
P(J_{ij})=\sqrt{\frac{N}{2 \pi {\tilde J}^2} }
   \exp \Bigl(-\frac{N}{2{\tilde J}^2} J_{ij}^2 \Bigr).
\]

We use the replica method to carry out the sample average 
\begin{eqnarray}
\Bigl[\bigl\langle AB \bigr\rangle \Bigr]&=&\lim_{n \to 0}
 \biggl[ \sum_{\mib \sigma}
A({\mib \sigma}^1, \J) B({\mib \sigma}^1, \J) 
\nonumber
\\
& &{} \times \eX \biggr]
\label{fir-term}
\\
\Bigl[\bigl\langle A \bigr\rangle \bigl\langle B \bigr\rangle \Bigr]
&=&\lim_{n \to 0} \biggl[ \sum_{\mib \sigma}
A({\mib \sigma}^1, \J) B({\mib \sigma}^2, \J) 
\nonumber\\
& &{} \times \eX \biggr],
\label{sec-term}
\end{eqnarray}
where replica indices $(\alpha=1,2,\cdots,n)$ are introduced.

It is necessary to calculate the average of various powers of
exchange interactions to derive the expressions 
of covariances in $\S4$.
It is straightforward to show by integration that
\begin{eqnarray}
& &{} \biggl[ J_{ab} \eXab \biggr] 
\nonumber\\
&=& \frac{{\tilde J}^2}{N} \beta \sum_\alpha \saa \sba \eYab
\label{n1}
\end{eqnarray}
and
\begin{eqnarray}
& &{}\biggl[ J_{ab}^2 \eXab \biggr] 
\nonumber\\
&=&\frac{{\tilde J}^2}{N} \eYab,
\label{n2}
\end{eqnarray}
where we have omitted terms which vanish in the $n \to 0$ limit.
Higher powers of the interaction lead to contributions with higher-power 
dependence on $1/N$.
That is, for $k\geq3$,
\begin{eqnarray}
& &{} \left|\biggl[ J_{ab}^k \eXab \biggr] \right|
\nonumber\\
&\leq& \frac{3{\tilde J}^4 \beta}{N^2} \left| \sum_\alpha \saa \sba \eYab\right| .
\label{n3}
\end{eqnarray}
Therefore we can neglect such higher-order terms
in the limit $N \to \infty$.

Let us now work on Cov($H \hh$) as an example.
Explicitly written,
\begin{eqnarray}
& &{}\mbox{Cov}\Bigl(H \hh \Bigr)=\lim_{n \to 0} \Biggl(
     \biggl[ \sum_{{\mib \sigma}}
     \Bigl(-\frac{1}{2} \sum_{ab} J_{ab} \sigma_a^1 \sigma_b^1 \Bigr) 
\nonumber\\
& \times& \Bigl(\sum_{cde} J_{cd} J_{de} \sigma_c^1 \sigma_e^1 \Bigr) 
      \eX \biggr]
\nonumber
\\
&-&\biggl[\sum_{{\mib \sigma}} 
\Bigl(-\frac{1}{2} \sum_{ab} J_{ab} \sigma_a^1 \sigma_b^1 \Bigr) 
\Bigl(\sum_{cde} J_{cd} J_{de} \sigma_c^2 \sigma_e^2 \Bigr)
\nonumber\\
&\times& \eX \biggr]\Biggr).
\label{H-hh}
\end{eqnarray}
It is straightforward to check that we need only terms 
in which there are overlaps between two interactions.
The two quantities $H$ and $\hh$ have four such cases,
i.e., $ab=cd,~ab=dc,~ab=de$ and $ab=ed$.
{}From the symmetry under the exchange of site indices,
these four cases give identical contributions
\begin{eqnarray*}
& &{}\mbox{Cov}\Bigl(H \hh\Bigr)=
\\
& &{}\lim_{n \to 0} 4\Biggl( -\frac{1}{2}\sum_{{\mib \sigma}}
\biggl[ \sum_{abe} J_{ab}^2 J_{be} \sigma_b^1 \sigma_e^1 \eX \biggr]
\\
& &{}+\frac{1}{2} \sum_{{\mib \sigma}} 
\biggl[ \sum_{abe} J_{ab}^2 J_{be} \sigma_a^1 \sigma_b^1 
\sigma_a^2 \sigma_e^2 \eX \biggr] \Biggr).
\end{eqnarray*}
The averages over $J_{ab}^2$ and $J_{be}$ are carried out independently.
Using eqs.~(\ref{n1}) and (\ref{n2}), the above equation is found to be
\begin{eqnarray*}
&& \lim_{n \to 0} \Biggl(-2 \sum_{abe}
\Bigl( \frac{{\tilde J}^2}{N}\Bigr) \Bigl( \frac{{\tilde J}^2}{N} \beta \Bigr)
\sum_{{\mib \sigma}} 
\biggl(\sum_\alpha \sba \sea \sigma_b^1 \sigma_e^1-
\\
&& \sum_\alpha \sba \sea \sigma_a^1 \sigma_b^1 \sigma_a^2 \sigma_e^2
\biggr) \eY \Biggr).
\end{eqnarray*}
The expression in the exponent has been rearranged so that
the problem reduces 
to that of a single-site system after the Hubbard-Stratonovitch
transformation.
The result of spin trace is, in the paramagnetic phase,
\begin{eqnarray}
\mbox{Cov}\Bigl(H \hh\Bigr)&=&-2 N{\tilde J}^4 \beta \biggl(
\sum_\alpha \delta_{\alpha 1} -\sum_\alpha \delta_{\alpha 1}
\delta_{\alpha 2} \delta_{1 2} \biggr)
\nonumber\\
&=&-2N{\tilde J}^4 \beta.
\end{eqnarray}
The other covariances can be calculated in the same way.
\section{Spin-field distribution function}
The dynamical spin-field distribution function
is expressed by the characteristic function defined in eq.~(\ref{lfd3}).
In this Appendix, we give detailed calculations of this function.

The method of high-temperature series expansion developed in $\S 2$ is
applicable to the evaluation of the characteristic function
\begin{equation}
\label{aplfd1}
G(k,\sigma)=\left[\frac{1}{N}\biggl<\sum_{i}\frac{1+\sigma\sigma_i}{2}
            \exp\Bigl(ik\sum_{j}J_{ij}\sigma_{j}\Bigr)\biggr>\right] .
\end{equation}

The first-order approximation in $\beta$ is obtained in the ordinary
way used in equilibrium calculations 
with  $\beta$ replaced by $\beta_{\rm eff}$.
This procedure is almost the same as the study by Thomsen {\it et al.}
\cite{ref:MT}
except for the factor $(1+\sigma\sigma_i)/2$

\begin{equation}
\label{aplfd2}
G^{(1)}(k,\sigma)=
  \left[\frac{1}{N}\biggl<\sum_{i}\frac{1+\sigma\sigma_i}{2}
  \exp\Bigl(ik\sum_{j}J_{ij}\sigma_{j}\Bigr)\biggr>_1\right] .
\end{equation}

We use the replica method and carry out the
average with respect to the distribution of $\{J_{ij}\}$,
so that the function is written as
\begin{full}
\begin{eqnarray}
&& G^{(1)}(k,\sigma)=\lim_{n\to0}\left[\frac{1}{N}{\sums}
   \sum_{i}\frac{1+\sigma\sigma_i^1}{2}
   \exp\Bigl( ik\sum_{j}J_{ij}\sigma_{j}^1 \Bigr) 
              \exp\Bigl( \beta_{\rm eff}\sum_{(jl)}J_{jl}u_{jl} \Bigr)\right]
\nonumber
\\
&=&\lim_{n\to0}\Biggl\{\frac{1}{N}{\sums}
   \sum_{i}\frac{1+\sigma\sigma_i^1}{2}
\exp\Bigl(\frac{{\tilde J}^2\beta_{\rm eff}^2}{2N}\sum_{(jl)}u_{jl}^2
   +\frac{ik{\tilde J}^2\beta_{\rm eff}}{N}\sum_lu_{il}\sigma_l^1
   -\frac{k^2{\tilde J}^2}{2}\Bigr)\Biggr\}
\nonumber
\\
&=&\lim_{n\to0}\Biggl\{\frac{1}{N}{\sums}
   \sum_{i}\frac{1+\sigma\sigma_i^1}{2}
     \exp\biggl(\frac{{\tilde J}^2nN\beta_{\rm eff}^2}{4}-\frac{k^2{\tilde J}^2}{2}
+ik{\tilde J}^2\beta_{\rm eff}\sigma_i^1
   +\frac{{\tilde J}^2\beta_{\rm eff}^2}{2N}\sum_{(\alpha\gamma)\neq(\alpha1)}
   \Bigl(\sum_l\sigma_l^{\alpha}\sigma_l^{\gamma}\Bigr)^2
\nonumber\\
&& +\frac{{\tilde J}^2\beta_{\rm eff}^2}{2N}\sum_{(\alpha1)}
   \Bigl(\sum_l\sigma_l^{\alpha}\sigma_l^1
    +\frac{ik\sigma_i^{\alpha}}{\beta_{\rm eff}}\Bigr)^2
   +\frac{k^2{\tilde J}^2}{N}\biggr)\Biggr\} ,
\label{aplfd5}
\end{eqnarray}
\end{full}
where $u_{ab}=\sum_{\alpha=1}^n\sigma_a^{\alpha}\sigma_b^{\alpha}$.
Application of the Hubbard-Stratonovitch transformation leads to
\begin{full}
\begin{eqnarray}
&& G^{(1)}(k,\sigma)=\lim_{n\to0}\Biggl\{
   \exp\left(\frac{{\tilde J}^2nN\beta_{\rm eff}^2}{4}-\frac{k^2{\tilde J}^2}{2}
    \Bigl(1-\frac{2}{N}\Bigr)\right)
 \int\biggl\{\prod_{(\alpha\gamma)}\Bigl(\sqrt{\frac{N}{2\pi}}
   {\tilde J}\beta_{\rm eff}\Bigr) {\rm d}q_{\alpha\gamma}\biggr\}
   \exp\Bigl(-\frac{N^2{\tilde J}^2\beta_{\rm eff}^2}{2}\sum_{(\alpha\gamma)}
   q_{\alpha\gamma}^2\Bigr)
   \nonumber\\
&& \times {\sums}\frac{1}{N}\sum_i\frac{1+\sigma\sigma_i^1}{2}
   \exp\biggl({\tilde J}^2\beta_{\rm eff}^2\sum_{(\alpha\gamma)\neq(\alpha1)}q_{\alpha\gamma}
   \sum_l\sigma_l^{\alpha}\sigma_l^{\gamma}
  \nonumber\\
&&+ {\tilde J}^2\beta_{\rm eff}^2\sum_{(\alpha1)}q_{\alpha1}
   \Bigl(\sum_l\sigma_l^{\alpha}\sigma_l^1
   +\frac{ik\sigma_i^{\alpha}}{\beta_{\rm eff}}\Bigr)
   +ik{\tilde J}^2\beta_{\rm eff}\sigma_i^1\biggr)\Biggr\} .
\label{aplfd6}
\end{eqnarray}
\end{full}
We evaluate the integrals at the saddle point of $q_{\alpha\gamma}$:
\begin{full}
\begin{eqnarray}
&& G^{(1)}(k,\sigma)
=\lim_{n\to0}\Biggl\{
   \exp\biggl(\frac{{\tilde J}^2nN\beta_{\rm eff}^2}{4}-\frac{k^2{\tilde J}^2}{2}
   \Bigl(1-\frac{2}{N}\Bigr)\biggr)
 \Bigl\{\sqrt{\frac{N}{2\pi}{\tilde J}\beta_{\rm eff}}\Bigr\}^{n(n-1)/2}
\exp\Bigl(-\frac{N^2{\tilde J}^2\beta_{\rm eff}^2}{2}
            \sum_{(\alpha\gamma)}q_{\alpha\gamma}^2\Bigr)
\nonumber\\
&& \times   {\sums}\frac{1}{N}\sum_i\frac{1+\sigma\sigma_i^1}{2}
   \exp\biggl({\tilde J}^2\beta_{\rm eff}^2\sum_{(\alpha\gamma)\neq(\alpha1)}q_{\alpha\gamma}
   \sum_l\sigma_l^{\alpha}\sigma_l^{\gamma}
     \nonumber\\
&&+{\tilde J}^2\beta_{\rm eff}^2\sum_{(\alpha1)}q_{\alpha1}
   \Bigl(\sum_l\sigma_l^{\alpha}\sigma_l^1
   +\frac{ik\sigma_i^{\alpha}}{\beta_{\rm eff}}\Bigr)
   +ik{\tilde J}^2\beta_{\rm eff}\sigma_i^1\biggr)\Biggr\} .
\label{aplfd7}
\end{eqnarray}
\end{full}
Furthermore, we assume the system in the paramagnetic phase, that is
$q_{\alpha\gamma}=\delta_{\alpha\gamma}$.
Finally, we obtain the first-order characteristic function as
\begin{eqnarray}
&& G^{(1)}(k,\sigma)\nonumber\\
\label{aplfd8}
&=&\lim_{n\to0}\Biggl\{\exp\Bigl(-\frac{k^2{\tilde J}^2}{2}\Bigr)
   \Bigl\{\sqrt{\frac{N}{2\pi}}{\tilde J}\beta_{\rm eff} \Bigr\}^{n(n-1)/2}\nonumber\\
& \times& {\sums}\frac{1}{N}\sum_i\frac{1+\sigma\sigma_i^1}{2}
   \exp\bigl(ik{\tilde J}^2\beta_{\rm eff}\sigma_i^1\bigr)\Biggr\}
\nonumber\\
&=&\exp\Bigl(-\frac{k^2{\tilde J}^2}{2}\Bigr)
   \frac{1}{2}\bigl(\cos(k{\tilde J}^2\beta_{\rm eff})
    \nonumber\\
   &+&i\sigma\sin(k{\tilde J}^2\beta_{\rm eff})\bigr) .
\end{eqnarray}

The higher order approximation needs covariances between
$(1/N)\sum_{i}(1+\sigma\sigma_i)/2\,\exp(ik\sum_{j}J_{ij}\sigma_{j})$ and
$A(\sigma,\{J_{ij}\})$, where $A$ stands for the quantities
appearing in the expansion of the dynamical probability distribution function.
The difference from the calculations of other quantities
is the existence of ($ik\sum_{j}J_{ij}\sigma_{j}$) in the exponent.
For this reason, the following term is included in equations:
\begin{eqnarray}
\lefteqn{\left[J_{ab}^l\exp\Bigl(\beta_{\rm eff} J_{ab}\sum_{(ij)}u_{ab}
         +ik\sum_cJ_{ac}\sigma_c^2\Bigr)\right]}\nonumber\\
\label{aplfd10}
&=&\Biggl[J_{ab}^l\exp\bigl(\beta_{\rm eff} J_{ab}u_{ab}
   +ikJ_{ab}\sigma_b^2\bigr)\Biggr]
\nonumber\\
&\times& \left[\exp\Bigl(\beta_{\rm eff} J_{ab}\sum_{(ij)\neq(ab)}u_{ab}
   +ik\sum_{c\neq b}J_{ac}\sigma_c^2\Bigr)\right] ,
\end{eqnarray}
where the coefficient $J_{ab}^l$ comes from the quantity $A(\sigma,\{J_{ij}\})$.
The index $l$ is smaller than or equal to three in this expansion,
because the quantity $A$ contains $J_{ab}^3$ at most.

The following relations are useful for sample averages:
\begin{eqnarray}
&& \Bigl[\exp\left(\beta_{\rm eff} J_{ab}u_{ab}+ikJ_{ab}\sigma_b^2\right)\Bigr]
\nonumber\\
\label{aplfd11}
&=&\exp\left(\frac{{\tilde J}^2}{2N}(\beta_{\rm eff}^2u_{ab}^2
   +2ik\beta_{\rm eff}\sigma_b^1u_{ab}-k^2)\right)
\end{eqnarray}
\begin{eqnarray}
&& \Bigl[J_{ab}\exp\left(\beta_{\rm eff} J_{ab}u_{ab}+ikJ_{ab}\sigma_b^2\right)\Bigr]
\nonumber\\
\label{aplfd12}
&=&\frac{{\tilde J}^2}{N}\left(\beta_{\rm eff} u_{ab}+ik\sigma_b^1\right)
\nonumber\\
&& \times \exp\left(\frac{{\tilde J}^2}{2N}(\beta_{\rm eff}^2u_{ab}^2
     +2ik\beta_{\rm eff}\sigma_b^1u_{ab}-k^2)\right)
\end{eqnarray}
\begin{eqnarray}
&& \Bigl[J_{ab}^2\exp\left(\beta_{\rm eff} J_{ab}u_{ab}+ikJ_{ab}\sigma_b^2\right)\Bigr]
\nonumber\\
\label{aplfd13}
&=&\left(\frac{{\tilde J}^2}{N}+\frac{{\tilde J}^4}{N^2}
    \left(\beta_{\rm eff} u_{ab}+ik\sigma_b^1\right)^2\right)
\nonumber\\
&& \times \exp\left(\frac{{\tilde J}^2}{2N}(\beta_{\rm eff}^2u_{ab}^2
       +2ik\beta_{\rm eff}\sigma_b^1u_{ab}-k^2)\right) .
\end{eqnarray}
In the case of $l=3$, the leading term is of order $1/N^2$.
Therefore there is no contribution in the thermodynamic limit
and we pay attention only to the case of $l\leq2$.

We explain a part of the calculation of the second-order approximation.
The second-order term is expressed as
\begin{eqnarray}
&& G^{(2)}(k,\sigma)=\beta^2b_1{\rm Cov}\biggl(\sum_{i}h_i^2,
   \sum_{i}\frac{1+\sigma\sigma_i}{2}
\nonumber\\
&& \times   \exp\Bigl(ik\sum_{j}J_{ij}\sigma_{j}\Bigr)\biggr)
\nonumber
\\
\label{aplfd15}
&=&\beta^2b_1\lim_{n\to0}\Biggl
 \{\biggl[\frac{1}{N}{\sums}\sum_{abcd}J_{ab}J_{ac}
   \sigma_b^1\sigma_c^1 
\nonumber\\
&& \times \exp\Bigl(ik\sum_eJ_{de}\sigma_e^1
   +\beta_{\rm eff}\sum_{(ij)}J_{ij}u_{ij}\Bigr)\biggr]
\nonumber\\
 && -\biggl[\frac{1}{N}{\sums}\sum_{abcd}J_{ab}J_{ac}\sigma_b^1\sigma_c^1 
\nonumber\\
 && \times \exp\Bigl(ik\sum_eJ_{de}\sigma_e^2
    +\beta_{\rm eff}\sum_{(ij)}J_{ij}u_{ij}\Bigr)\Biggr] .
 \label{aplfd155}
\end{eqnarray}
The overlap between the interactions makes the covariance non-vanishing,
but this does not mean that all overlaps give finite contributions.
In the present case, three configurations $d=a,b$ and $c$
have non-vanishing covariances
and others are zero.

Let us consider the configuration $d=a$ as an example.
The difference between the first and the second terms
is only in the replica index of $\sigma_e$ in the exponent.
Thus we explain only the procedure to treat the second term.
The procedure is similar to the equilibrium calculation\cite{ref:MT}
and the second term is found to be expressed as
\begin{full}
\begin{eqnarray}
&& \beta^2b_1
   \lim_{n\to0}\Biggl\{\biggl[\frac{1}{N}{\sums}\sum_{abc}J_{ab}J_{ac}
   \sigma_b^1\sigma_c^1
 \exp\Bigl(ik\sum_eJ_{ae}\sigma_e^2+\beta_{\rm eff}
   \sum_{(ij)}J_{ij}u_{ij}\Bigr)\biggr]\nonumber
\\
&=&\beta^2b_1\lim_{n\to0}\Biggl\{\frac{1}{N}
    {\sums}\sum_{abc}\sigma_b^1\sigma_c^1
  \biggl[J_{ab}\exp\Bigl(ik\sum_eJ_{ab}\sigma_e^2
     +\beta_{\rm eff} J_{ab}u_{ab}\Bigr)\biggr]
\biggl[J_{ac}\exp\Bigl(ik\sum_eJ_{ac}\sigma_e^2 
+\beta_{\rm eff} J_{ab}u_{ac}\Bigr)\biggr]
   \nonumber\\
&& \times \biggl[\exp\Bigl(ik\sum_{e\neq b,c}J_{ae}\sigma_e^2
   +\beta_{\rm eff}\sum_{(ij)\neq(ab),(ac)}J_{ij}u_{ij}\Bigr)\biggr]\Biggr\}
\label{aplfd16}
\\
&=&\beta^2b_1\lim_{n\to0}\Biggl\{\frac{{\tilde J}^4}{N^3}{\sums}\sum_{abc}\sigma_b^1
   \sigma_c^1(\beta_{\rm eff} u_{ab}+ik\sigma_b^2)(\beta_{\rm eff} u_{ac}+ik\sigma_c^2)
\nonumber\\
&& \times\exp\biggl(\frac{{\tilde J}^2}{2N}\Bigl(\sum_{(ij)}
   \beta_{\rm eff}^2u_{ij}^2+\sum_e2ik\beta_{\rm eff}
   \sigma_e^2u_{ae}\Bigr)-\frac{k^2{\tilde J}^2}{2}\biggr)\Biggr\} .
\label{aplfd17}
\end{eqnarray}
\end{full}
Applying the Hubbard-Stratonovitch transformation and the
saddle-point method,
we obtain in the paramagnetic phase
\begin{eqnarray}
\label{aplfd18}
&& \beta^2b_1\frac{{\tilde J}^4}{N^3}{\sums}\sum_{abc}\biggl\{
   \beta_{\rm eff}^2\sum_{\gamma\delta}\sigma_a^{\gamma}\sigma_a^{\delta}\sigma_b^1
   \sigma_b^{\gamma}\sigma_c^1\sigma_c^{\delta}
\nonumber\\
&&  +ik\beta_{\rm eff}\Bigl(\sum_{\gamma}\sigma_a^{\gamma}\sigma_b^1\sigma_b^{\gamma}
   \sigma_c^1\sigma_c^2
 +\sum_{\delta}\sigma_a^{\delta}\sigma_b^1\sigma_b^2
   \sigma_c^1\sigma_c^{\delta}\Bigr)
\nonumber\\
&&   -k^2\sigma_b^1\sigma_b^2\sigma_c^1\sigma_c^2\biggr\}
   \exp\left(-\frac{k^2{\tilde J}^2}{2}+ik{\tilde J}^2\beta_{\rm eff}\sigma_a^2\right) .
\end{eqnarray}
By tracing out the spin configurations, we find the only 
non-vanishing contribution in the first term for the case
$\gamma=\delta=1$.  In this way
the second term of eq.~(\ref{aplfd155}) is written as
\begin{equation}
\label{aplfd19}
\beta^2b_1\left\{{\tilde J}^4\beta_{\rm eff}^2\exp\left(-\frac{k^2{\tilde J}^2}{2}
\cos(k{\tilde J}^2\beta_{\rm eff})\right)\right\} ,
\end{equation}
in the limit of $N\to\infty$.

The first term is obtained by the exchange $\sigma^2 \to \sigma^1$ as
\begin{eqnarray}
\label{aplfd20}
&& \beta^2b_1\frac{{\tilde J}^4}{N^3}{\sums}\sum_{abc}\Biggl\{
   \beta_{\rm eff}^2\sum_{\gamma\delta}\sigma_a^{\gamma}\sigma_a^{\delta}\sigma_b^1
   \sigma_b^{\gamma}\sigma_c^1\sigma_c^{\delta}
\nonumber\\
&&   +ik\beta_{\rm eff} \Bigl(\sum_{\gamma}\sigma_a^{\gamma}\sigma_b^1\sigma_b^{\gamma}
   \sigma_c^1\sigma_c^1
 +ik\beta_{\rm eff}\sum_{\delta}\sigma_a^{\delta}\sigma_b^1\sigma_b^1
   \sigma_c^1\sigma_c^{\delta} \Bigr)
\nonumber\\
&&   -k^2\sigma_b^1\sigma_b^1\sigma_c^1\sigma_c^1\Biggr\}
   \exp\left(-\frac{k^2{\tilde J}^2}{2}+ik{\tilde J}^2\beta_{\rm eff}\sigma_a^1\right) .
\end{eqnarray}
Non-vanishing contributions come from terms satisfying $\gamma=\delta=1$:
\begin{eqnarray}
\label{aplfd21}
&& \beta^2b_1\Biggr\{{\tilde J}^4\beta_{\rm eff}^2\exp
   \left(-\frac{k^2{\tilde J}^2}{2}\right)\cos(k{\tilde J}^2\beta_{\rm eff})
\nonumber\\
&& -2k{\tilde J}^4\beta_{\rm eff}\exp\left(-\frac{k^2{\tilde J}^2}{2}\right)\sin(k{\tilde J}^2\beta_{\rm eff})
\nonumber\\
&& -k^2{\tilde J}^4\exp\left(-\frac{k^2{\tilde J}^2}{2}\right)\sin(k{\tilde J}^2\beta_{\rm eff})\Biggr\} .
\end{eqnarray}
Subtraction of eq.~(\ref{aplfd19}) from eq.~(\ref{aplfd21}) gives
the covariance of the configuration $d=a$ in eq.~(\ref{aplfd16}) as
\begin{eqnarray}
\label{aplfd22}
&& \beta^2b_1\Biggl\{-2k{\tilde J}^4\beta_{\rm eff}\exp
    \left(-\frac{k^2{\tilde J}^2}{2}\right)\sin(k{\tilde J}^2\beta_{\rm eff})
\nonumber\\
&& -k^2{\tilde J}^4\exp\left(-\frac{k^2{\tilde J}^2}{2}\right)\sin(k{\tilde J}^2\beta_{\rm eff})\Biggr\} .
\end{eqnarray}

For the configurations of $d=b$ and $c$,
the covariances are identical and the result is
\begin{equation}
\label{aplfd23}
\beta^2b_1\left\{-k{\tilde J}^4\beta_{\rm eff}\exp\left(-\frac{k^2{\tilde J}^2}{2}\right)
  \sin(k{\tilde J}^2\beta_{\rm eff})\right\} .
\end{equation}
Contributions of all configurations are summed up to yield
\begin{eqnarray}
\label{aplfd24}
G^{(2)}(k,\sigma) &=&
   \beta^2b_1\Biggl\{-4k{\tilde J}^4\beta_{\rm eff}\exp
   \left(-\frac{k^2{\tilde J}^2}{2}\right)\sin(k{\tilde J}^2\beta_{\rm eff})
   \nonumber\\
&& -k^2{\tilde J}^4\exp\left(-\frac{k^2{\tilde J}^2}{2}\right)\sin(k{\tilde J}^2\beta_{\rm eff})\Biggr\} .
\end{eqnarray}

The third-order term can be calculated similarly.
%

\newpage
{\bf Figure Captions}
\vspace{0.1in}

Fig. 1.  The relaxation of $[\langle H \rangle]/N$ of the SK model by 
Monte Carlo simulation, high-temperature expansions, and the
approximation of a Boltzmann form with
the inverse temperature $\beta_{\rm eff}^{(3)}$.
Average and standard deviation are shown for $50$ samples simulated.
The system size is $N=1000$ in simulations.
\vspace{0.1in}

Fig. 2.  The relaxation of the expectation value of $\sum h_i^2$ per spin 
for the  SK model.
\vspace{0.1in}

Fig. 3.  The relaxation of the expectation value of $\sum \sigma_i h_i^3$ 
per spin for the SK model.
\vspace{0.1in}

Fig. 4.  The relaxation of the expectation value of $\sum J_{ij} h_i h_j$ 
per spin for the SK model.
\vspace{0.1in}

Fig. 5.  The relaxation of $[\langle H \rangle]/N$ of the
two-dimensional $\pm J$ model by Monte Carlo simulation,
the first order approximation and
the $\beta_{\rm eff}^{(3)}$-approximation.
Average and standard deviation are shown for $50$ samples simulated.
The system size is $N=50^2$ in simulations.
\vspace{0.1in}

Fig. 6.  The relaxation of the expectation value of $\sum h_i^2$ per spin 
for the two-dimensional $\pm J$ model.
\vspace{0.1in}

Fig. 7.  The relaxation of the expectation value of $\sum \sigma_i h_i^3$ 
per spin for the two-dimensional $\pm J$ model.
\vspace{0.1in}

Fig. 8.  The relaxation of the expectation value of $\sum J_{ij} h_i h_j$ 
per spin for the two-dimensional $\pm J$ model.
\vspace{0.1in}

Fig. 9.  Time evolution of the spin-field distribution
functions $P(h,\pm1)$ at $T=2$.
Circles and squares are results of simulations.
Solid lines denote the theoretical prediction of the third-order 
approximation.
\vspace{0.1in}

Fig. 10.  Time evolution of physical quantities
by the third-order series expansions (full curves)
and simulations (dotted curves, standard deviations omitted,
$N=1000$ and $50$ samples)
when the SK model is quenched from $T_0=\infty$ to $T=2$.
\vspace{0.1in}

Fig. 11.  Time evolution of correlation coefficients with $\sum J_{ij} h_i h_j$
when the system is quenched from $T_0=\infty$ to $T=2$
obtained by using $\beta_{\rm eff}^{(3)}$.
\vspace{0.1in}

Fig. 12.  Time evolution of the terms in the dynamical 
probability distribution function evaluated by the third-order series 
expansions for the SK model quenched from $T_0=\infty$ to $T=5$.
\vspace{0.1in}

Fig. 13.  Time evolution of the terms 
in the dynamical probability distribution function
evaluated by the third-order series-expansions
for the SK model quenched from $T_0=\infty$ to $T=2$.
\vspace{0.1in}

Fig. 14.  Time evolution of the terms 
in the dynamical probability distribution function
for the SK model quenched from $T_0=\infty$ to $T=1=T_{\rm c}$ by simulations.
The system size is $N=1000$.
The results are $50$-sample averaged.
\vspace{0.1in}

Fig. 15.  Time evolution of the terms 
in the dynamical probability distribution function
for the two-dimensional 
$\pm J$ model quenched from $T_0=\infty$ to $T=5$
by the $\beta_{\rm eff}^{(3)}$-approximations.
\vspace{0.1in}

Fig. 16.  Time evolution of the terms 
in the dynamical probability distribution function
for the two-dimensional
$\pm J$ model quenched from $T_0=\infty$ to $T=2$ by simulations.
The system size is $N=50^2$ and
the results are averaged over $50$ samples.
\vspace{0.1in}

Fig. 17.  First-order cumulants of the spin-field distribution
functions $P(h,\pm1)$ at $T=2$.
The dotted line is by the high-temperature expansion and
the full line is for the CLS theory.  Simulation results
($N=5000$, 100 samples) are also shown for comparison.
\vspace{0.1in}

Fig. 18.  Second-order cumulants of the spin-field distribution
functions $P(h,\pm1)$ at $T=2$.
Symbols are the same as in Fig. 17.


\end{document}